%% file: main.tex
\documentclass[
    preprints,
    article,
    accept,
    pdftex,
    moreauthors]{Definitions/mdpi}

\usepackage{multirow}
\usepackage{amsmath}
\usepackage{booktabs}
\usepackage{graphicx}
\usepackage{tabularx}
\usepackage{adjustbox}
\usepackage{lscape}
\usepackage{siunitx}

\graphicspath{{figures/}}

\newcommand{\Nsft}{N_{\mathrm{SFT}}}
\newcommand{\Tsft}{T_{\mathrm{SFT}}}
\newcommand{\Sn}{S_{\mathrm{n}}}

\newcommand{\barP}{\bar{P}}
\newcommand{\D}{\mathcal{D}}
\newcommand{\pd}{p_{\mathrm{V}}}
\newcommand{\kurt}{\kappa}
\newcommand{\kurth}{\kurt_{\mathrm{th}}}

\firstpage{1} 
\pubvolume{1}
\issuenum{1}
\articlenumber{0}
\pubyear{2023}
\copyrightyear{2023}
\datereceived{ } 
\daterevised{ } 
\dateaccepted{ } 
\datepublished{ } 
\hreflink{https://doi.org/} 

\Title{Assessing the similarity of continuous gravitational-wave signals to narrow instrumental artifacts}

\TitleCitation{Assessing the similarity of continuous gravitational-wave signals to narrow instrumental artifacts}

\Author{Rafel Jaume $^{*}$\orcidA{}, Rodrigo Tenorio\orcidB{} and Alicia M. Sintes\orcidC{}}

\AuthorNames{Rafel Jaume, Rodrigo Tenorio and Alicia M. Sintes}

\AuthorCitation{Jaume, R.; Tenorio, R.; Sintes, A. M.}

\address{%
Departament de Física, Institut d’Aplicacions Computacionals i de Codi Comunitari (IAC3),\\
Universitat de les Illes Balears, and Institut d’Estudis Espacials de Catalunya (IEEC),\\
Carretera de Valldemossa km 7.5, E-07122 Palma, Spain
}

\corres{Correspondence: rafel.jaume@uib.es}

\abstract{
    Continuous gravitational-wave signals (CWs) are long-lasting quasi-monochromatic gravitational-wave signals
    expected to be emitted by rapidly-rotating non-axisymmetric neutron stars. Depending on the rotational
    frequency and sky location of the source, certain CW signals may behave in a similar manner to
    narrow-band artifacts present in ground-based interferometric detectors. 
    Part of the detector-characterisation tasks in the current generation of interferometric detectors
    (Advanced LIGO, Advanced Virgo, and KAGRA) aim at understanding the origin of these narrow artifacts,
    commonly known as ``spectral lines''. It is expected that similar tasks will continue after the arrival of
    next-generation detectors (e.g. Einstein Telescope and Cosmic Explorer).
    Typically, a fraction of the observed lines in a given
    detector can be associated to one or more  instrumental causes; others, however, have an unknown origin. 
    In this work, we assess the similarity of CW signals to spectral lines in order to understand whether
    a CW may be mistaken for a noise artifact. 
   Albeit astrophysically unlikely, our results
   do not rule out the possibility of a CW signal being visible in the detector's power spectrum.
}


\begin{document}

\section{Introduction}
\label{sec:introduction}
\input{introduction}

\section{Characterizing the visibility of continuous-wave signals}
\label{sec:introduce_kurtosis}
\input{introduce_kurtosis}

\section{Implications for astrophysical CW sources}
\label{sec:expected_signals}
\input{expected_signals}

\section{Conclusion}
\label{sec:conclusion}
\input{conclusion}

\authorcontributions{All authors have read and agreed to the published version of the manuscript.}

\funding{
    This work was supported by the
    Universitat de les Illes Balears (UIB); the Spanish Agencia
    Estatal de Investigación grants PID2022-138626NB-I00,
    PID2019-106416GB-I00, RED2022-134204-E, RED2022-
    134411-T, funded by MCIN/AEI/10.13039/501100011033;
    the MCIN with funding from the European Union
    NextGenerationEU/PRTR (PRTR-C17.I1); Comunitat
    Autònoma de les Illes Balears through the Direcció General
    de Recerca, Innovació I Transformació Digital with funds
    from the Tourist Stay Tax Law(PDR2020/11 - ITS2017-
    006), the Conselleria d’Economia, Hisenda i Innovació
    co-financed by the European Union and FEDER Operational
    Program 2021-2027 of the Balearic Islands; the “ERDF A
    way of making Europe”.
    R.~J.~is supported by the Conselleria de Educació,
    Universitat i Recerca del Govern de les Illes Balears FPI/018/2020.
}

\acknowledgments{
    We thank Ansel Neunzert, Evan Goetz, David Keitel, and the Continuous Waves working group of the LIGO-Virgo-KAGRA Collaboration
    for useful comments and discussions. This paper has been assigned document number LIGO-P2300463.
    }

\conflictsofinterest{The authors declare no conflict of interest.}

\begin{adjustwidth}{-\extralength}{0cm}

\reftitle{References}


\externalbibliography{yes}
\bibliography{references.bib}

\end{adjustwidth}
\end{document}

%% file: introduction.tex
Continuous gravitational waves (CWs) are long lasting
gravitational wave signals (GW) whose detection remains,
so far, unattained~\cite{Riles:2022wwz,Wette:2023dom}.
Amongst the expected sources we find rapidly-spinning non-axisymmetric
neutron stars (NS)~\cite{Sieniawska:2019hmd}, but also other more exotic
ones such as evaporating boson clouds formed around spinning
black holes~\cite{Zhu:2020tht,Jones:2023fzz},
or planetary-mass compact binary systems~\cite{Miller:2020kmv,Miller:2021knj}.
Their expected frequency lies on the audible band of the GW spectrum, which
makes them prime targets to be detected using the current generation of advanced
ground-based interferometric detectors (Advanced LIGO~\cite{LIGOScientific:2014pky},
Advanced Virgo~\cite{VIRGO:2014yos} and KAGRA~\cite{KAGRA:2018plz}),
as well as third-generation detectors (3G) (Einstein Telescope (ET)~\cite{Maggiore:2019uih}
and Cosmic Explorer (CE)~\cite{Reitze:2019iox}).

The frequency evolution of a CW signal as emitted by a rapidly-spinning NS can
usually be described using a Taylor series~\cite{Jaranowski:1998qm}
\begin{equation}
    f_{\mathrm{s}}(t) = f_0 + (t - t_{\mathrm{ref}}) f_1 + \dots \,,
    \label{eq:f_s}
\end{equation}
where $f_0$ is the initial CW frequency at a fiducial reference time $t_\mathrm{ref}$
and $f_{k \geq 1}$ are the \emph{spindown} parameters. 
The number of terms to include in Eq.~\eqref{eq:f_s} is generally dependent on the age
of the source: older objects tend to spin down more slowly and thus require a lower number of
terms to be accurately modelled~\cite{Krishnan:2004sv,Knispel:2008ue,Wade:2012qc}.
Typical searches for unknown CW sources include one or two spindown terms in order to remain
computationally affordable~\cite{Tenorio:2021wmz}. Searches for younger objects,
such as newborn neutron stars, try to detect CW emission on a much shorter timescale 
(hours to days) and tend to be conducted using the general torque equation as discussed
in~\cite{Sarin:2018vsi, Oliver:2019ksl, Grace:2023kqq}.

Upon arrival to the detector, the frequency of a CW signal is Doppler-modulated due
to the detector's motion around the Solar System Barycenter (SSB)
\begin{equation}
    f(t) = f_{\mathrm{s}}(t)  \cdot \left(1 + \frac{\vec{v}(t)}{c} \cdot \hat{n} \right)\,,
    \label{eq:doppler}
\end{equation}
where $c$ is the speed of light, $\hat{n}$ is the sky position of the source in the SSB,
and $\vec{v}$ is the velocity of the detector with respect to the SSB.
The detector's velocity can be further expanded into two components
$\vec{v} = \vec{v}_{\mathrm{o}} + \vec{v}_{\mathrm{r}}$,
where $v_{\mathrm{o}} \approx 10^{-4} c$ is the Earth's orbital speed
and $v_{\mathrm{r}} \approx 10^{-6} c$ is the Earth's rotational speed.
This translates into two time-dependent and sky-location-dependent frequency modulations.
The magnitude of the Doppler modulation is proportional to $\vec{v} \cdot \hat{n}$.
Since the time-averaged orientation of $\vec{v}$ is close to the ecliptic plane,
sky positions near the poles  (for which $\vec{v} \cdot \hat{n} \approx 0$)
are expected to suffer the smallest frequency modulation~\cite{Prix:2005gx, Leaci_2015}.

Ground-based interferometric detectors are affected by a wide variety
of noise sources that behave in a similar manner to GW signals~\cite{Berger:2018ckp}.
For example, short noise transients (``glitches'')  are known to overlap
with GWs produced in the coalescense of two compact
objects~\cite{Cabero:2019orq,LIGO:2021ppb, Vazsonyi:2022jul, Davis:2022ird}.
If unaddressed, noise artifacts significantly degrade the sensitivity of a GW search,
in the sense that astrophysically interestiong candidates will likely be recovered at a higher
false alarm probability (i.e. \emph{lower} significance level).

CW searches are mainly affected by spectrally narrow and persistent noise artifacts,
also known as ``lines'' due to their line-like appearance in a power spectrum~\cite{PaperPep}.
Lines are initially identified by inspecting the power spectral density of a detector,
either manually or using a peak-finding algorithm, and then further classified into
``instrumental lines'' if they can be associated to an instrumental cause,
or ``unknown lines'' if otherwise. 
Well-known instrumental lines include, for example, lines at multiples of the $\SI{60}{\hertz}$
electrical power frequency in the Advanced LIGO detectors or at multiples of the the resonance
frequency of the mirror suspensions (about $\SI{500}{\hertz}$).
Lists of instrumental and unknown lines in the third observing run of the
Advanced LIGO detectors (O3) can be found in~\cite{Identified, Unidentified}, respectively.

Lines tend to cause a high number of outliers in CW searches, as their relatively strong power
and persistence with respect to the detector's background PSD causes them to look like
excess power for an analysis.
This is specially problematic in broad searches for unknown sources,
such as all-sky searches, as their CW models tend to be particularly sensitive to lines
due to the use of low coherence times~\cite{Tenorio:2021wmz}\footnote{
    A notable exception are the \texttt{Einstein@Home} CW searches
    (see~\cite{Steltner:2023cfk} and references therein), which use a detection statistic
    that extends the noise hypothesis to include lines~\cite{Keitel:2013wga}.
}. To deal with such an elevated number of candidates, CW searches follow two main approaches:
First, CW candidates can be tested using a broad suite of consistency vetoes in order to
find an anomalous behavior in their amplitude or frequency
evolution~\cite{Leaci_2015, Zhu:2017ujz,Valluri:2020cqe,Tenorio:2021njf,Jones:2022fgs}.
Second, the frequency evolution of a candidate can be cross-checked against a list
of~\emph{instrumental} lines in order to understand whether its significance
is caused by the crossing of one or several lines.

Moderately-strong CW signals may appear as as narrow features in the power spectrum
(see Fig.~\ref{fig:averaged_power}). These features will become more obvious as the sensitivity
of the interferometric detectors improves~\cite{Valluri:2020cqe}. Even for the current generation
of detectors, the possibility of a CW signal being visible in the detector has not been ruled out.
Vetoing candidates near narrow spectral features, without additional evidence of their terrestrial
origin, would result in an as-yet unquantified increase in the false dismissal probability of true CW signals.

Current practices in the LVK collaboration therefore discourage the use of~\emph{unknown} lines
lists to veto CW candidates. The presence of a visible artifact in the data without
a clear instrumental origin is not considered enough evidence for a candidate to be deemed
non-astrophysical. However, No systematic study exists, to date, to back up such a recommendation.

In this paper, we study the similarity of CW signals and spectral lines in the current and next generation
of interferometric detectors. Specifically, we construct a statistical criterion to quantify whether a CW
signal is ``visible'' in the power spectrum of an interferometric detector. This criterion will be used to 
understand if artifacts in detector data can be consistent with astrophysical CW signals and strengthen
the recommendation of not using unidentified lines to veto the results of a CW search for unknown sources.
The paper is structured as follows: in Sec.~\ref{sec:introduce_kurtosis} we introduce basic statistical notation
and a criterion to quantify the  ``visibility'' of a CW signal in a power spectrum.
In Sec.~\ref{sec:expected_signals}, we compare the distribution of visible CW signals to an optimistic 
astrophysical distribution of sources. Conclusions are drawn in Sec.~\ref{sec:conclusion}.

%% file: introduce_kurtosis.tex
In this section, we construct a quantitative criterion to flag a CW signal as ``visible''
in a power spectrum. A visible signal would be susceptible of being flagged as an
unknown line. This criterion will be used in Sec.~\ref{sec:expected_signals} to understand
whether astrophysically-possible CW sources may produce visible signals
and understand the severity of using unknown lines in a search veto procedure.

The output of a ground-based GW detector can be described as a time series
of additive zero-mean Gaussian noise $n$ and, possibly, a GW signal $h$
\begin{equation}
    x = n + h \,.
\end{equation}
CW analyses usually make use of short-time Fourier transforms (SFTs)~\cite{SFT},
which are the Fourier transforms of short data segment with a duration $\Tsft$, typically
less than a few hours:
\begin{equation}
   \tilde{x}^{\alpha}(f) = \Delta t \sum_{m=0}^{M-1} \,  x^{\alpha}_m e^{-2 \pi i m \Delta t f} \,.
\end{equation} 
Here, the superscript ${\alpha}$ refers to the time at the beginning of an SFT, $M$ is the 
number of data samples within an SFT, and  $\Delta t = \Tsft / M$. Frequency resolution is
related to SFT duration as $\delta f = \Tsft^{-1}$.
Noise within an SFT can be assumed to be white and stationary due to their short duration.
As a result, the noise is fully characterized by its single-sided power spectral density (PSD) $\Sn$.

The optimal strategy to identify a monochromatic signal in Gaussian noise is to identify
local maxima in the frequency-domain spectrum~\cite{bretthorst}. This strategy remains useful
for quasi-monochromatic signals as long as the frequency modulations are small compared to the
frequency resolution of the dataset and the noise distribution remains stationary.
For the case of long-duration narrow-band signals, non-stationarities in the noise distribution 
can be dealt with by normalizing the power in each SFT according to their PSD 
\begin{equation}
    P^{\alpha}(f) = \frac{4}{\Tsft \Sn^{\alpha}(f) } \left| \tilde{x}^{\alpha}(f) \right|^2 \,,
    \label{eq:normalized_power}
\end{equation}
where we used the well-known relation between variance and PSD~\cite{Krishnan:2004sv}
\begin{equation} 
    \langle \left| \tilde{n}^{\alpha} \right|^2 \rangle = \frac{1}{2} \Tsft \Sn^{\alpha}.
\end{equation}
The collection of time-frequency normalized-power values $P^{\alpha}(f)$ is usually referred
to as a ``spectrogram''. This spectrogram can then be averaged along time (i.e.~SFTs)
to reveal the presence of persistent signals: 
\begin{equation}
    \barP(f) = \frac{1}{\Nsft} \sum_{\alpha=1}^{\Nsft} P^{\alpha}(f) \,.
    \label{eq:averaged_power}
\end{equation}
We shall refer to the frequency-dependent quantity $\barP(f)$ as ``power spectrum''.
Line identification tasks consists in identifying narrow artifacts in $\barP(f)$ and
cross-correlating them with an instrumental cause in the detector~\cite{PaperPep}. 

The statistical properties of $\barP$ are well-known.
Assuming Gaussian noise, $P^{\alpha}$ is the sum of the squares of two
zero-mean unitary-variance Gaussian random variables, thus it follows chi-squared
distribution with two degrees of freedom $P^{\alpha} \sim \chi^2_{2}$.
$\barP$, on the other hand, is the average of $\Nsft$ identical and independent random variables,
with usually $\Nsft \sim \mathcal{O}(10^{3} - 10^{4})$. Thus, due to the central limit theorem,
$\barP$ follows a Gaussian distribution with mean $\mu = 2$ and standard deviation $\sigma = 2$, 
$\barP \sim \mathrm{Gauss}(2, 2)$. As a result, \emph{the presence of artifacts in the data
is related to deviations from Gaussianity in the distribution of $\barP$.}

We will assume throughout this work that data consist of Gaussian noise and a single CW signal.
Thus, the visibility of a CW signal is directly related to how much $\barP$ deviates from Gaussianity.
The characterization of a similar method to be applied as a decision criterion on a real-data situation
is beyond the scope of this work and left for future work. 

Tests for deviations from Gaussianity are commonly referred to as ``normality tests''~\cite{normality_tests}.
These have customarily been used in a CW searches to identify parameter-space regions contaminated by
noise artifacts~\cite{Tenorio:2021wmz}. In order to choose a specific normality test,
one needs to understand what kind of deviations from Gaussianity are expected in the data.
For instance, Ref.~\cite{Owen:2023maw} uses the \mbox{Cramér-von Mises}  statistic~\cite{cramer},
to flag overpopulations of outliers within a few standard deviations of the mean; 
this is motivated by the fact that CW signals, on the other hand, tend to populate the very
far end of a distribution's tail and thus have a negligible contribution to the Cramér-von Mises
statistic .

\begin{figure}
    \includegraphics[width=0.8\textwidth]{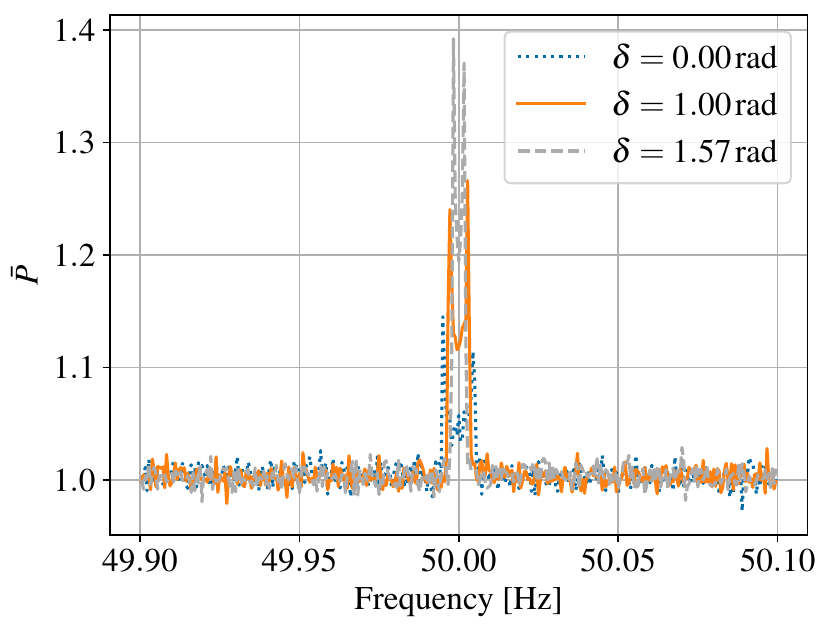}
    \includegraphics[width=0.8\textwidth]{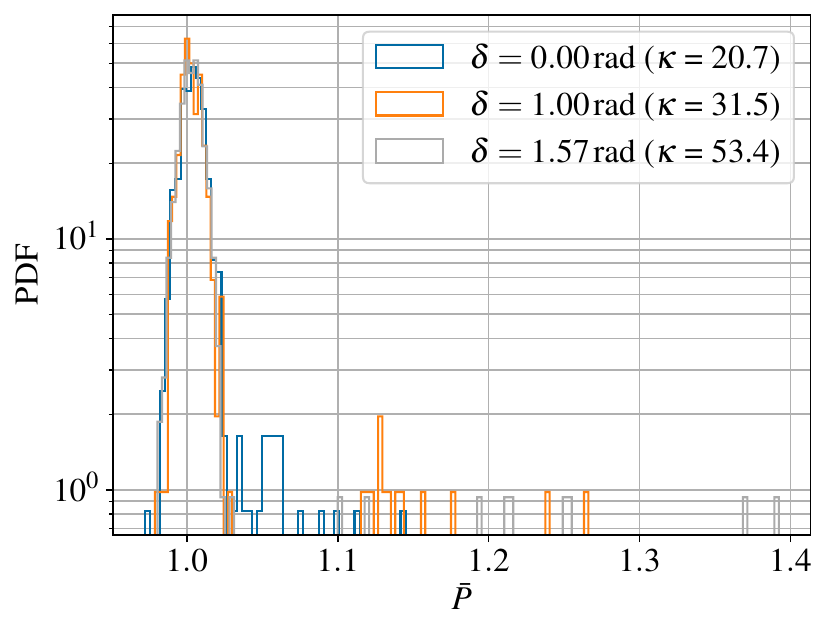}
    \caption{
        (Top) Power spectra for the three CW signals at $f_0 = \SI{50}{\hertz}$ and null
        spindown parameter.
        (Bottom) Distribution of power spectra and kurtosis for the three same CW signals.
        As discussed in Sec.~\ref{sec:introduction}, the width of the spectral artifact is related to
        the sky position of the CW source. Sources away from the sky poles 
        (low declination $\left| \delta \right| \lesssim 1$) tend to produced broader signals
        than sources closer to the sky poles (high declination $\left| \delta \right| \approx 1.5$),
        The amplitudes of these signals corresponds to a sensitivity depth of $\mathcal{D} = 15$. 
        for illustration purposes. We refer the reader to Sec.~\ref{sec:kurt_from_cw}
        and Sec.~\ref{sec:expected_signals} for a detailed discussion on relevant CW
        amplitudes.
    }
    \label{fig:averaged_power}
\end{figure}

In Fig.~\ref{fig:averaged_power}, we show three power spectra containing CW signals at
different sky locations. The behavior displayed by a CW signal is that of a relatively
narrow disturbance that significantly shifts the power of the affected frequency bins well into the
positive tail of the background noise distribution. As thoroughly discussed in~\cite{kurtosis},
the presence of such outliers can be readily measured by the excess kurtosis
\begin{equation}
    \kurt = \left\langle \left(\frac{\barP - \mu_{\barP}}{ \sigma_{\barP}}\right)^4  \right\rangle- 3 \,,
\end{equation}
where 
\begin{equation}
    \mu_{\barP} = \langle \barP \rangle \,,
\end{equation}
\begin{equation}
    \sigma_{\barP}^2 = \langle \barP ^2 \rangle - \langle \barP \rangle^2 \,,
\end{equation}
and angle brackets denote ensemble average. The excess kurtosis is constructed so that
$\kurt = 0$ for a Gaussian variable, since $\langle (X - \mu)^4\rangle = 3 \sigma^2$
for any Gaussian random variable $X$ with mean $\mu$ and standard deviation $\sigma$.
Kurtosis is expected to be positive if the distribution of $\barP$ contains outliers
significantly displaced from the background average, which makes it an appropriate
tool to construct a visibility criterion.

In the following subsections, we characterize $\kurt$ to quantify
the visibility of a CW signal in Gaussian noise. These results will then be used in
Sec.~\ref{sec:expected_signals} to assess the visibility of astrophysical CW signals.

\subsection{Expected kurtosis from a finite Gaussian sample}
\label{subsec:kurtosis_noise}

\begin{figure}
    \center
    \includegraphics[width=0.8\textwidth]{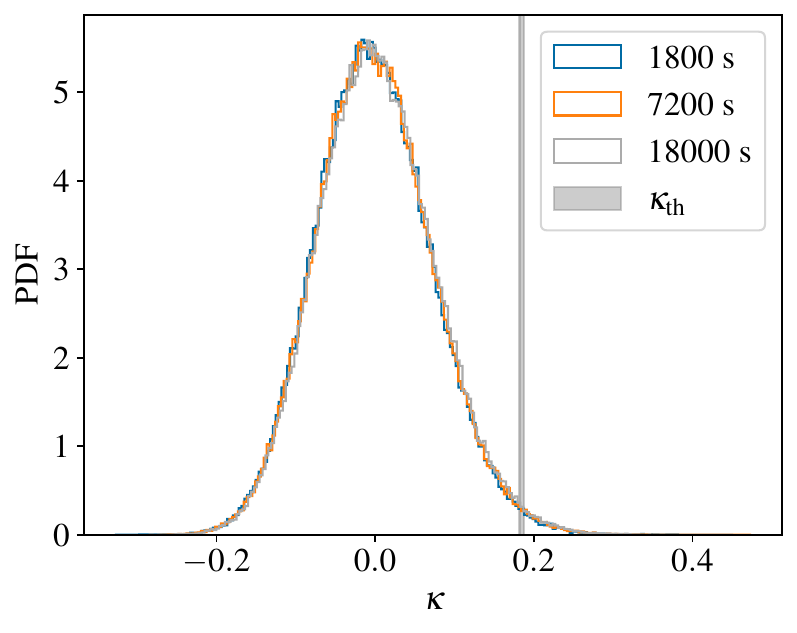}
    \caption{
        Distribution of sample kurtosis using $1.6 \cdot {10^5}$ Gaussian noise realizations
        for \mbox{$\Tsft = \SI{1800}{\second}, \SI{7200}{\second}, \SI{18000}{\second}$}.
        Each realization lasts for 1 year and contains 4500
        frequency bins. Thresholds corresponding to the 99\% quantile are highlighted in gray
        and collected in Table~\ref{table:thresholds}.
    }
    \label{fig:noise_kurtosis}
\end{figure}

Several kurtosis estimators have been proposed in the literature~\cite{sample_kurtosis, FisherKurtosis}.
The distribution of these estimators is generally dependent on the distribution of the underlying samples,
and must be characterized in order to cogently quantify deviations from Gaussianity.
In this work, we compute the sample kurtosis using the \texttt{scipy.stats.kurtosis} function as
implemented in \texttt{SciPy}~\cite{2020SciPy-NMeth}. To characterize kurtosis distribution,
shown in Fig.~\ref{fig:noise_kurtosis}, we numerically generate $1.6\cdot 10^5$ noise power spectra
and compute the sample kurtosis. 

Samples are generated using \texttt{lalpulsar\_Makefakedata\_v5}~\cite{lalsuite} to simulate
a 1-year Gaussian-noise datastream with a PSD of $\Sn = \SI{1e-46}{\hertz^{-1}}$. SFTs were generated
using three different time lengths, namely $\Tsft = \SI{1800}{\second}, \SI{7200}{\second}, \SI{18000}{\second}$.
For each SFT, we computed their power spectrum [Eq.~\eqref{eq:averaged_power}], which then was
used to compute the sample's kurtosis $\kurt$. The frequency band $\Delta f$ was adjusted according to $\Tsft$
so that all samples contained $\Delta f / \delta f = 4500$ frequency bins; as a result, sample kurtosis is always
estimated using the same number of samples and thus the kurtosis distributions for all $\Tsft$ values are
comparable.

\begin{figure}
    \center
    \includegraphics[width=0.81\textwidth]{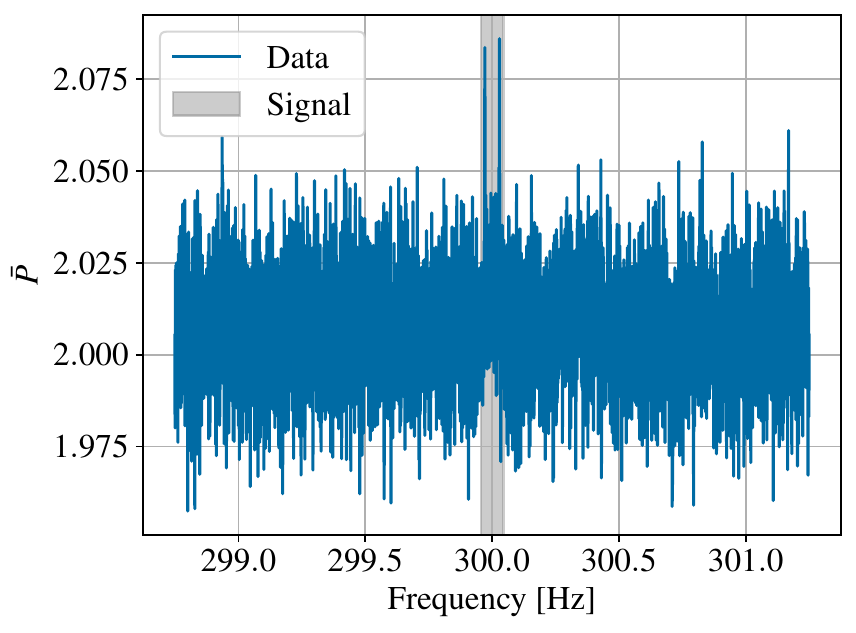}
    \includegraphics[width=0.8\textwidth]{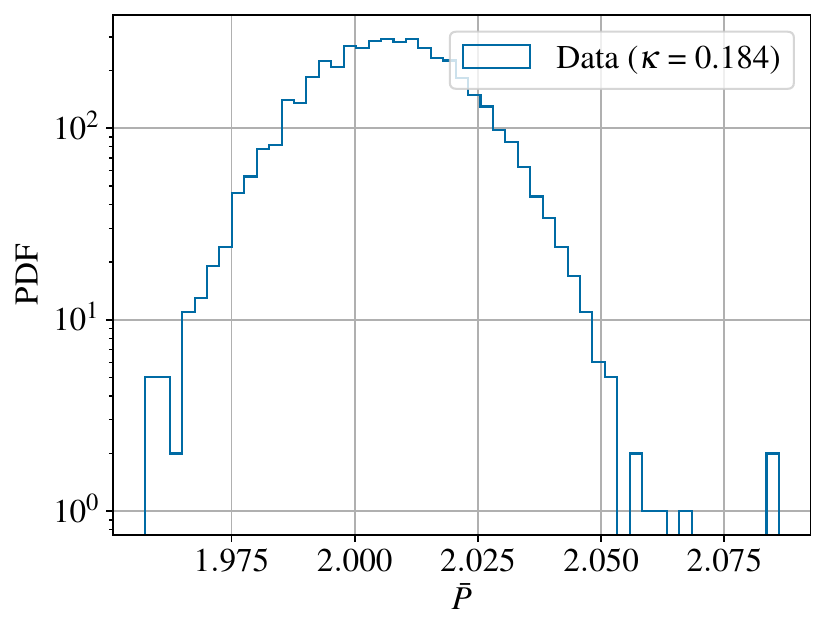}
    \caption{
        Power spectrum of the Advanced LIGO Hanford detector containing
        a ``visible'' CW signal according to the kurtosis criterion $\kurt > \kurth$. 
        Data was generated using $\Tsft = \SI{1800}{\second}$ and $\Sn = \SI{1e-46}{\hertz^{-1}}$.
        Signal parameters (see Sec.~\ref{sec:kurt_from_cw}) are $\mathcal{D} = 9$, 
        $\cos\iota = 0.450$, $\alpha = 2.653$, $\delta = 0$, $\phi=0.994$, $\psi = 0.178$.
    }
    \label{fig:visible_signal}
\end{figure}

We select a kurtosis threshold $\kurth$ corresponding to the 99\% quantile of the resulting kurtosis
distribution for each $\Tsft$, as shown in Table~\ref{table:thresholds}.
Any Gaussian-noise power spectrum yielding a kurtosis above $\kurth$ will be flagged as containing a visible CW signal.
We exemplify this visibility criterion in Fig.~\ref{fig:visible_signal}, where we show ``visible'' CW signal according to
our criterion $\kurt > \kurth$.

\begin{table}[]
    \center
    \begin{tabular}{cccc}
        \toprule
        $\Tsft$ &\SI{1800}{\second} & \SI{7200}{\second} & \SI{18000}{\second} \\
        \midrule
        $\kurth$ & 0.182 &0.185 & 0.188  \\
        \bottomrule
    \end{tabular}
    \caption{
        Kurtosis thresholds corresponding to the 99\% quantile of the numerically-generated
        kurtosis distribution shown in Fig.~\ref{fig:noise_kurtosis}. We note that lower $\Tsft$
        values tend to cause lower $\kurth$. Since the duration of the observing run is fixed to
        1 year, a lower $\Tsft$ value causes a higher number of SFTs to be generated. This causes
        the distribution of the resulting $\barP$ to be closer to a Gaussian, which in turn diminishes
        the spread of the kurtosis distribution. 
    }
    \label{table:thresholds}
\end{table}

This criterion will be used in Sec.~\ref{sec:kurt_from_cw} to compute the visible fraction of an
astrophysically-motivated population of CW sources. Concretely, we will compute the CW amplitude
at which a certain fraction of the population of CW signals returns a kurtosis above $\kurth$.
The choice of $\kurth$ as the 99\% quantile implies that the amplitudes corresponding to
small visible fractions (e.g. less than 10\% of visible signals) will be slightly overestimated,
as 1\% of the strictly non-visible signals will pass the threshold due to noise fluctuations. 
The results here reported, thus, will be conservative in the sense that the risks associated
to a CW signal being visible in the power spectrum will not be underestimated.

\subsection{Kurtosis from CW signals}
\label{sec:kurt_from_cw}

To characterize the visibility of CW signals in a power spectrum,
we simulate a population of CW sources with different amplitudes and determine the fraction
of signals above the kurtosis threshold $\kurth$ computed in Sec.~\ref{subsec:kurtosis_noise}.
We then compute, using a numerical fit, the amplitude at which a representative fraction of signals.
This process is akin to sensitivity estimation procedures used in CW searches
(e.g.~\cite{KAGRA:2022dwb} and references therein).

We simulate a population of all-sky isotropically oriented sources. This corresponds to sampling
the sky-position angles $(\alpha, \delta)$ from a uniform distribution on the sky sphere,
and uniformly sampling the polarisation angle $\psi \in \left[-\pi/4, \pi/4\right]$, the cosine
of the inclination angle $\cos \iota \in \left[-1, 1\right]$, and the initial phase
$\phi_0 \in \left[ 0, 2 \pi \right]$. 

We select 8 representative frequency bands across the frequency range expected for 
a CW signal, namely $\left[ 10, 30, 100, 200, 300, 500, 700, 1000\right]\,\unit{\hertz}$.
This choice is motivated by the fact that the bandwidth of the Doppler modulations induced
on the CW are proportional to the CW's frequency [Eq.~\eqref{eq:doppler}]:
higher frequencies are expected to modulate along broader frequency bands,
and thus produce less prominent peaks than lower frequencies.

For simplicity, we set all the spindown parameters to zero.
This is consistent with the usual spindown values observed in the known pulsar population~\cite{atnf}.
Non-zero spindown values would cause a broader modulation of the CW's frequency, which in turn
would diminish the prominence of the resulting peak in the power spectrum, reducing the visibility
of a signal. This choice is inline with producing a conservative result that does not underestimate
the visibility of a CW signal in the power spectrum.

The CW amplitude $h_0$ is specified in terms of the sensitivity depth
$\mathcal{D}$~\cite{Behnke:2014tma,Dreissigacker:2018afk,Wette:2023dom}
\begin{equation}
    \mathcal{D} = \frac{\sqrt{\Sn \, \mathrm{Hz}} }{h_0} \,.
    \label{eq:depth}
\end{equation}
We select a range of sensitivity depth values from 0.1 to 80 in order to sample the full range
of visibility fractions. PSD is fixed to $\Sn = \SI{1e-46}{\hertz^{-1}}$.

Throughout this analysis, we simulate signals as seen in a 1-year observing run of
the Advanced LIGO Hanford detector. Since 1) the duration of the observing run is on the
order of a year, 2) CW sources are uniformly distributed across the sky, 3) CW signals span a 
very narrow frequency band, in the sense that the PSD can be considered constant, and
4) amplitudes are expressed relative to PSD by means of the sensitivity depth, the resulting
sensitivity depths are valid for any other ground-based interferometric detector, and will
be used in Sec.~\ref{sec:kurt_from_cw} to estimate the visibility of a possible astrophysical
population of CW sources for different detectors.

\begin{figure}
    \center
    \includegraphics[width=0.7\textwidth]{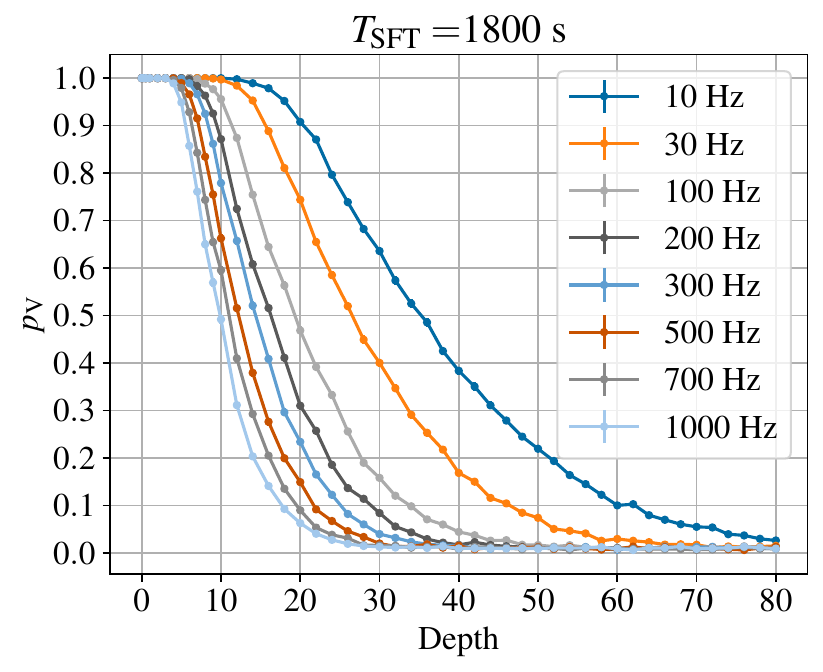}
    \includegraphics[width=0.7\textwidth]{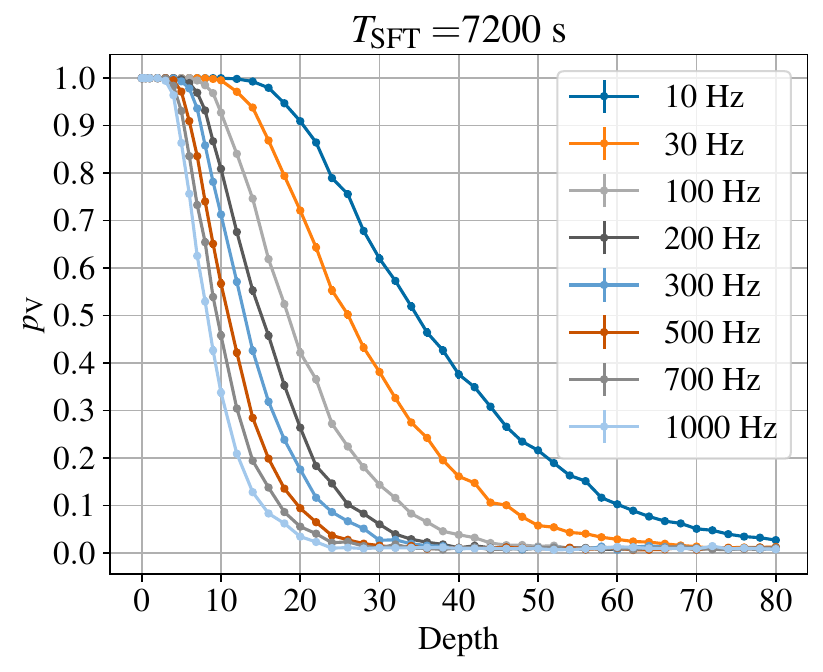}
    \includegraphics[width=0.7\textwidth]{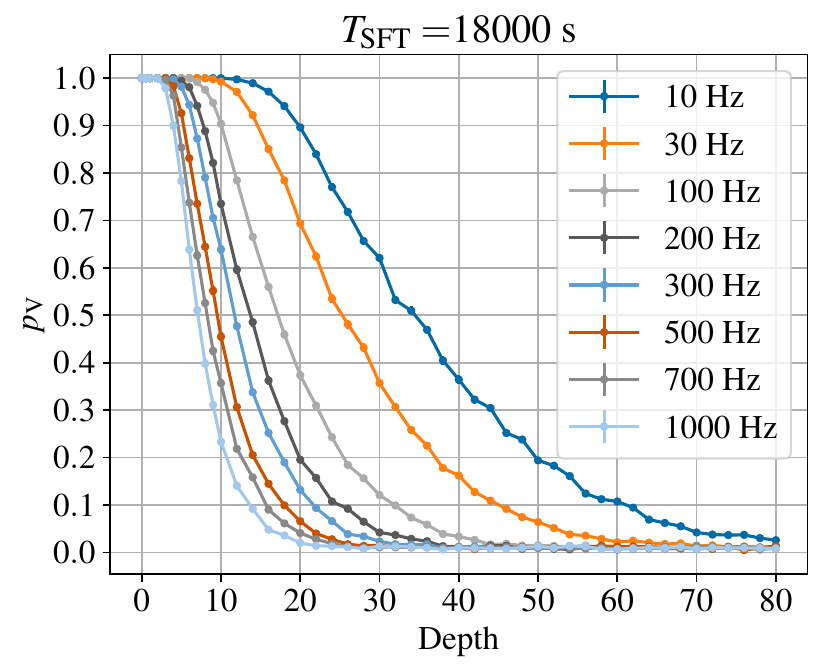}
    \caption{
        Fraction of visible signals $\pd$ for a population of all-sky isotropically-oriented
        signals at different frequencies and sensitivity depths. Each simulated signal is injected
        in a 1-year data set for the Advanced LIGO H1 detector. Each computed $\pd$ value
        has an associated binomial uncertainty of less than $1\%$.
        }
    \label{fig:pd}
\end{figure}

For each depth and frequency band, we simulate $N_{\mathrm{I}} = 3000$ CW signals
in Gaussian noise and compute the fraction of visible signals $\pd(\mathcal{D})$ as 
\begin{equation}
    \pd(\D) = \frac{1}{N_{\mathrm{I}}}  \sum_{n=1}^{N_{\mathrm{I}}} \begin{cases} 1, & \mathrm{if}\, \kurt^{(n)} > \kurth\\ 0, & \mathrm{otherwise}\end{cases} \,,
\end{equation}
where $\kurt^{(n)}$ is the kurtosis of the $n$-th simulated signal.
This quantity describes the fraction of visible signals in a population with a given constant depth.
More generally, $\pd$ is the probability of drawing a \emph{visible} signal from the specified
population at a given depth. The results are shown in Fig.~\ref{fig:pd}. 

The specific $\pd$ value of interest depends on the application at hand.
we fit a sigmoid-like ansatz using the function \texttt{NonlinearModelFit} as implemented in
\texttt{Mathematica}~\cite{Mathematica} to $\pd$
\begin{equation}
    \pd(\D; a, b) = \frac{1}{1 + b \D^{5/2} e^{-a/\D}} \,
\end{equation}
and compute the visible sensitivity depth at $\pd = 10\%$ and $\pd = 90\%$.

\begin{figure}
    \centering
    \includegraphics[width=0.8\textwidth]{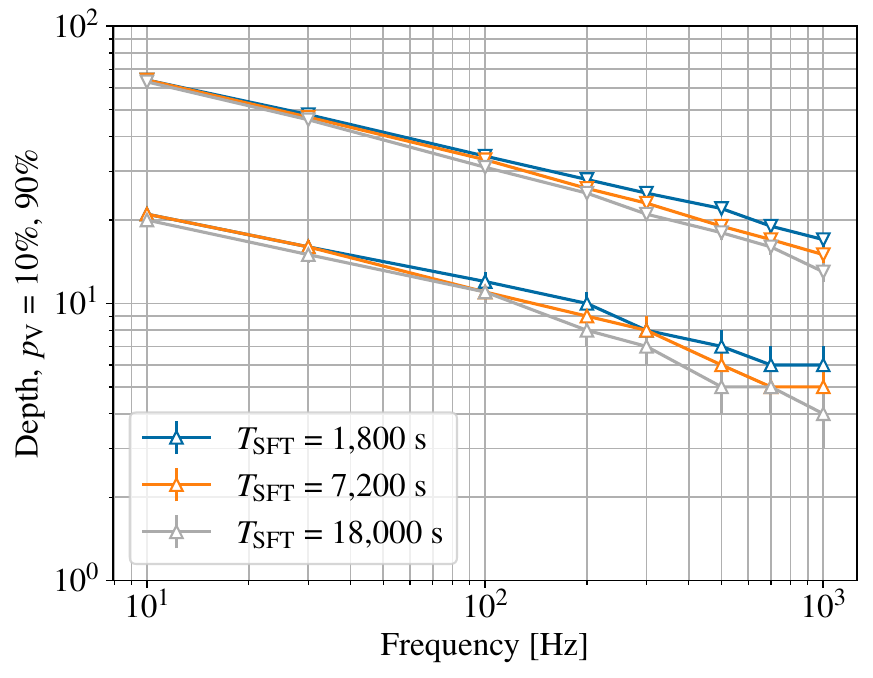}
    \caption{
        Sensitivity depth for $\pd = 10\%$ (downard triangles) and $\pd = 90\%$
        (upward triangles) for different frequencies as reported in Table~\ref{table:depths}.
    }
    \label{fig:depht}
\end{figure}

The resulting sensitivity depths are listed in Table~\ref{table:depths} and shown in Fig.~\ref{fig:depht},
and follow the expected trend: higher frequencies produce a broader Doppler modulation, and thus 
require a higher amplitude (lower depth) in order to produce a visible peak in the power spectrum.
The higher $\pd$, the bigger the fraction of visible signals: as a result, at a given frequency, 
$\mathcal{D}$ tends to \emph{decrease} (higher amplitudes) as we increase $\pd$. As discussed
in Table~\ref{table:depths}, the results appear to be shifted for different $\Tsft$ values. This
is a result of the different kurtosis thresholds, as discussed in Table~\ref{table:thresholds}.
From Fig.~\ref{fig:depht}, we can conclude that low-frequency CW signals tend to be
``more visible'' than high-frequency signals at a similar amplitude, in the sense that the lower
Doppler modulation allows for a more prominent peak to be displayed.

These sensitivity depths should be interpreted as the relative amplitude with respect
to the background noise required for a CW signal to be visible in a power spectrum. For a given
detector PSD, these values can be converted to the corresponding CW amplitude $h_0$ using
Eq.~\eqref{eq:depth}. To understand the risks associated to a visible CW signal,
such as mistakenly flagging it as an unknown line or whether visible artifacts are enough to 
ascribe a non-astrophysical origin, however, we should compare these results to the expected
amplitude from a population of astrophysical CW sources.

\begin{table}[]
    \center
    \begin{tabular}{cccccccccc}
        \toprule
        $\Tsft$ & $\pd$ & \SI{10}{\hertz} & \SI{30}{\hertz} & \SI{100}{\hertz} &\SI{200}{\hertz} &\SI{300}{\hertz} &\SI{500}{\hertz} &\SI{700}{\hertz} &\SI{1000}{\hertz} \\
        \midrule
        $\SI{1800}{\second}$
        & $90\%$ & 21 & 16 & 12 & 10 & 8 & 7 & 6 & 6\\
        & $10 \%$ & 64 & 48 & 34 & 28 & 25 & 22 & 19 & 17\\
        $\SI{7200}{\second}$
        & $90\%$ & 21 & 16 & 11 & 9 & 8 & 6 & 5 & 5 \\
        & $10 \%$ & 64 & 47 & 33 & 26 & 23 & 19 & 17 & 15 \\
        $\SI{18000}{\second}$
        & $90\%$ & 20 & 15 & 11 & 8 & 7 & 5 & 5 & 4\\
        & $10 \%$ & 63 & 46 & 31 & 25 & 21 & 18 & 16 & 13 \\
        \bottomrule
    \end{tabular}
    \caption{
        Sensitivity depth $\D$ for $\pd = 10\%$ and $\pd = 90\%$
        as estimated from the results shown in Fig.~\ref{fig:pd}.
        The uncertainty in each value, which we estimated by propagating the 
        binomial uncertainty on the empirical $\pd$ using the covariance matrix
        of the numerical fit, is $\pm 1$ for all the values in the table.
        Depth values appear to be systematically biased towards higher values
        as we reduced $\Tsft$. This is because $\kurth$ is lower for lower
        $\Tsft$ values, as discussed in Table~\ref{table:thresholds}. 
    }
    \label{table:depths}
\end{table}

%% file: expected_signals.tex
The visible sensitivity depths recorded in Table~\ref{table:depths} must be compared to a
realistic estimate of the expected amplitude of a CW signal in order to assess the risks of
associated to a visible CW signal in the power spectrum.
Since no CW has been detected to date, we will follow the conservative approach taken so far
and construct an optimistic astrophysical CW source (Sec.~\ref{subsec:optimistic_cw}) to compare
against the obtained results for different detector configurations (Sec.~\ref{subsec:ifos}).

\subsection{An optimistic CW source}
\label{subsec:optimistic_cw}

The nominal amplitude of a CW signal $h_0$ emitted by a NS with a
quadrupolar deformation parametrized by the equatorial ellipticity
$\varepsilon$ is given by~\cite{Jaranowski:1998qm}
\begin{equation}
    h_0 = \frac{4 \pi^2 G}{c^4} I f_0^2\frac{\varepsilon}{d} \,,
    \label{eq:h0}
\end{equation}
where  $f_0$ is the CW frequency (twice the rotational frequency in this model),
$d$ is the distance from the NS to the detector, and 
$I = 10^{38}\,\mathrm{kg}\,\mathrm{m}^2$ is the canonical moment
of inertia of a NS around the spinning axis~\cite{LIGOScientific:2007leh}\footnote{
    As discussed in~\cite{LIGOScientific:2007leh}, the moment of inertia differs
    by a factor of $\sim 3$ for different equations of state. Conversely, as discussed
    in the text, the equatorial ellipticity has a broader dynamical range depending on the
    sourcing mechanism; as a result, we choose to fix the moment of inertia to the canonical
    value and focus our discussion on $\varepsilon$.
}.
The results from this section could also be re-interpreted assuming other
other emission mechanisms,  such as r-modes or free
precession~\cite{Tenorio:2023hzh,Wette:2023dom}.

The latests theoretical bounds on the maximum ellipticity sustained by
a NS are $\varepsilon \sim 10^{-6}$~\cite{Gittins:2021zpv,Morales:2022wxs}.
This upper bound increases by a few orders of magnitude if more exotic
objects are considered~\cite{Owen:2005fn}. On the lower end, the observed
population of millisecond pulsars appears to sustain an ellipticity of
about $\varepsilon \sim 10^{-9}$~\cite{Woan:2018tey}.

\begin{figure}
    \center
    \includegraphics[width=0.7\textwidth]{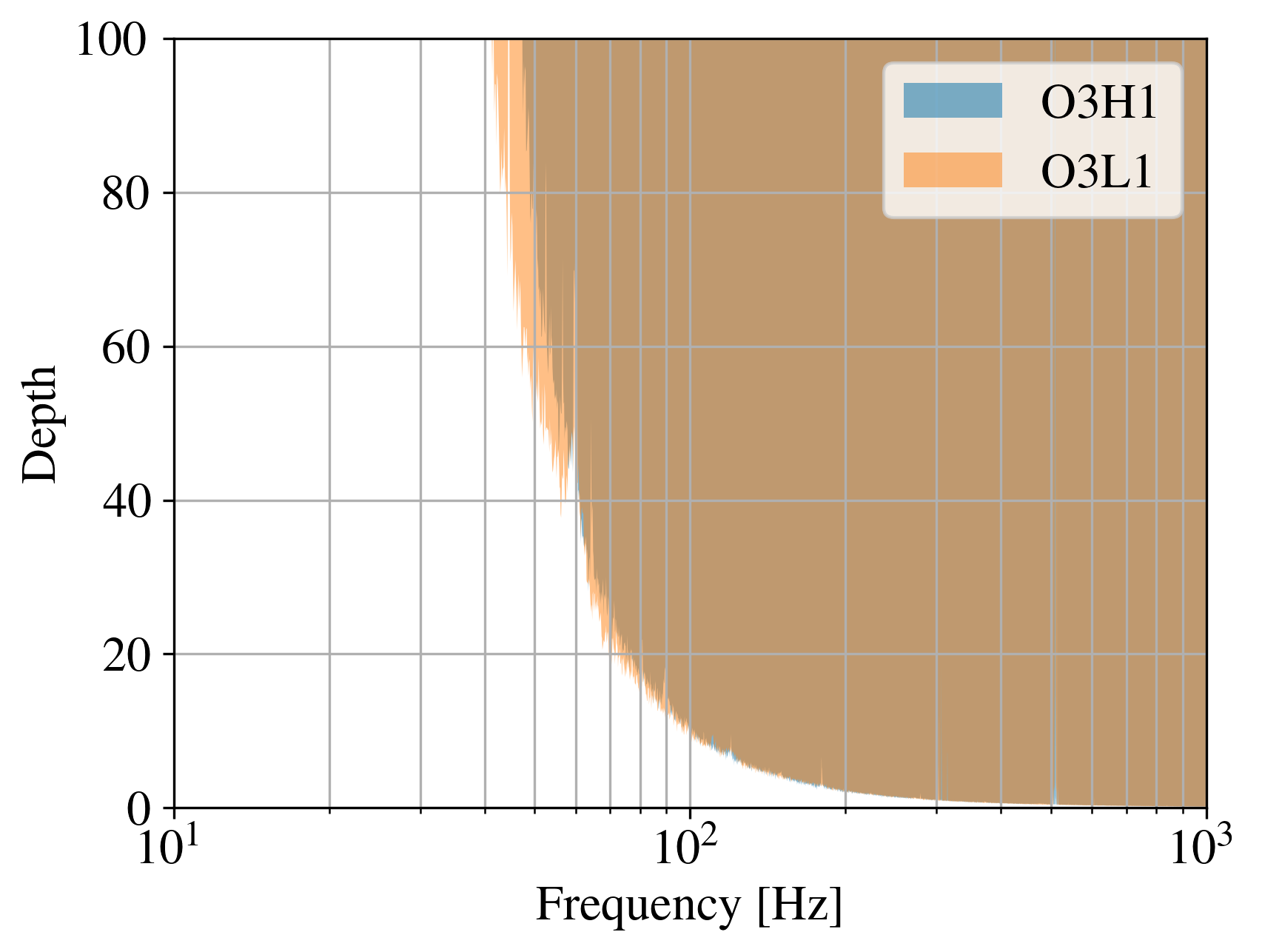}
    \includegraphics[width=0.7\textwidth]{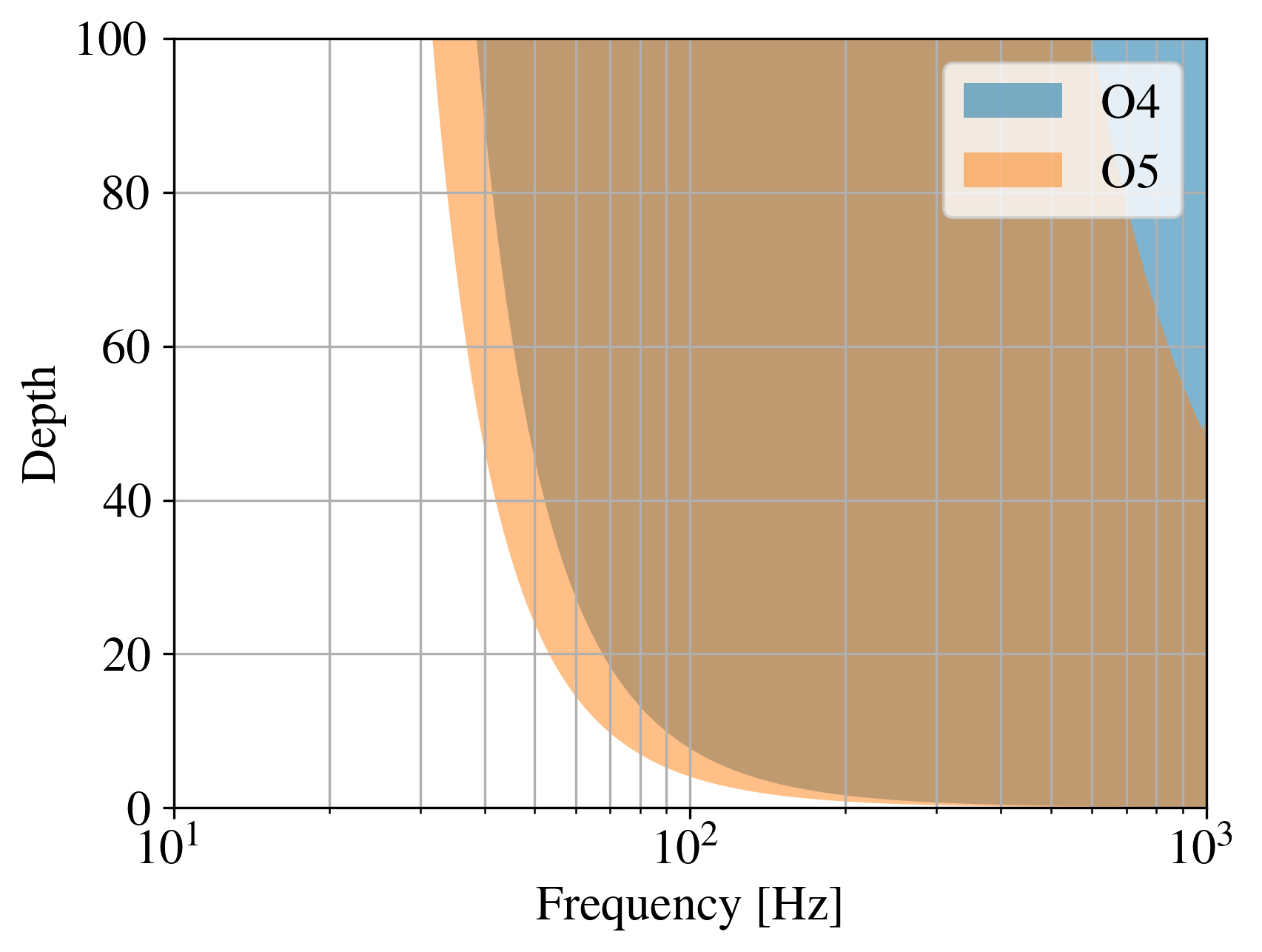}
    \includegraphics[width=0.7\textwidth]{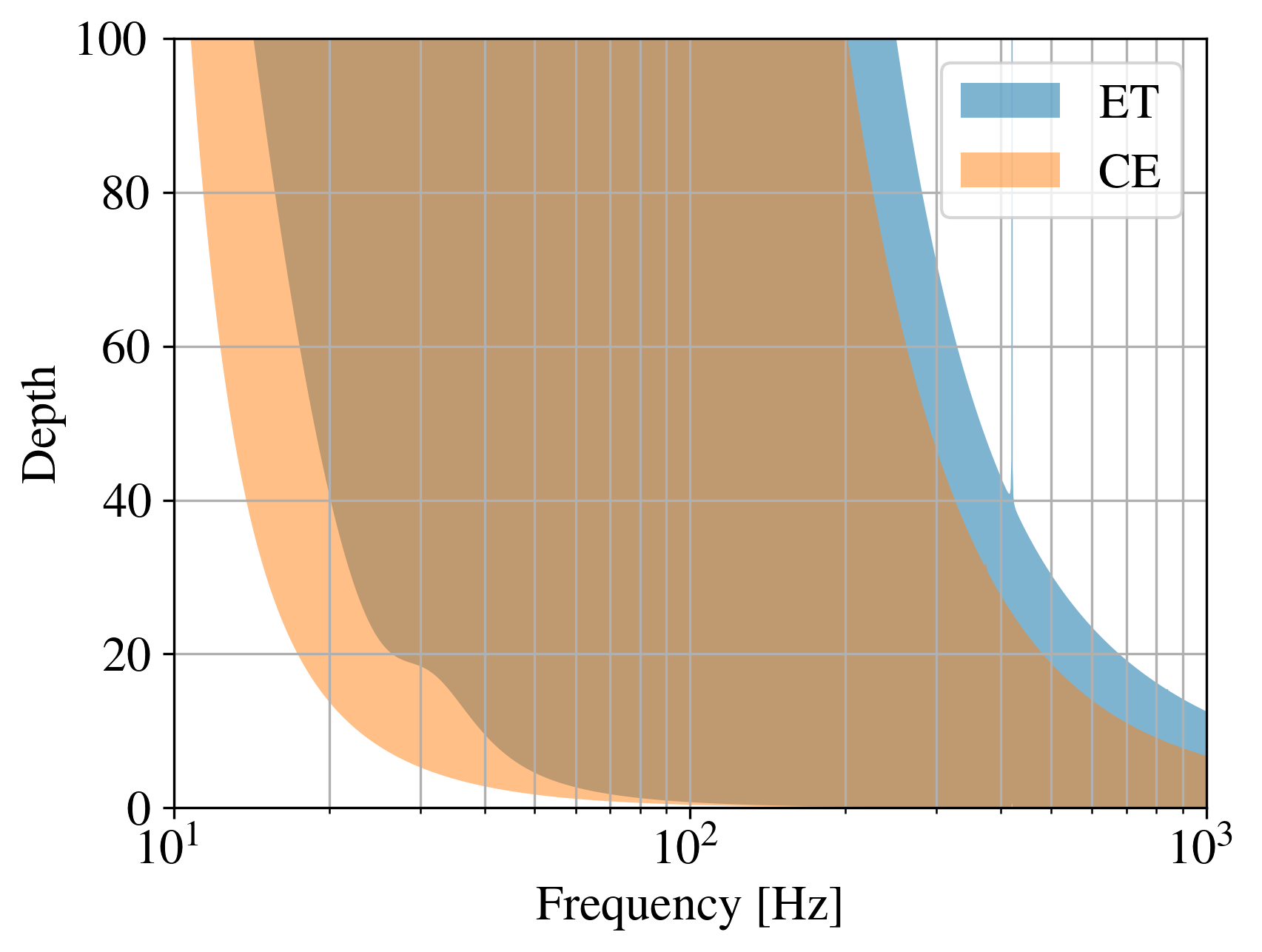}
    \caption{
        Sensitivity depth for an optimistic CW source located at $d = 20\,\mathrm{pc}$
        for different ground-based interferometric detectors.
        The lower limit of the shaded area corresponds to $\varepsilon = 10^{-6}$;
        the upper limit corresponds to $\varepsilon = 10^{-9}$. The presence of narrow 
        artifacts in the O3 sensitivity depths is due to the use a PSD estimated using
        real-data~\cite{O3bH1ASD,O3bL1ASD}. The shoulder-like plateau at low frequencies
        for the ET detector corresponds to the transition of the low-frequency interferometer
        to the high-frequency interferometer~\cite{Hild:2009ns}.
        }
    \label{fig:ellipticity_ranges}
\end{figure}

We construct an optimistic CW source by considering the full range of plausible
ellipticities $\varepsilon \in [10^{-9}, 10^{-6}]$ for a NS located at
$d = 20\,\mathrm{pc}$, which corresponds to the closest location to the detector
at which a NS is expected to be found~\cite{Sartore}.
Note that $h_0$ depends on the ratio $\varepsilon / d$; for example,
the expected amplitude of a source with
$\varepsilon = 10^{-6}$ at $2\,\mathrm{kpc}$ is equal to that of a source with
$\varepsilon = 10^{-8}$ at $20\,\mathrm{pc}$. In Fig.~\ref{fig:ellipticity_ranges}.
we show the expected sensitivity depth for optimistic sources
in the sensitive band of the Advanced LIGO
(O3, O4, and O5 sensitivities)~\cite{O3bH1ASD,O3bL1ASD,O4ASD,O5ASD},
Einstein Telescope~\cite{ETASD,Hild:2009ns}, and Cosmic Explorer~\cite{CEASD}
detectors. This is computed by using Eq.~\eqref{eq:h0} and the available PSD curves in the
literature.

The resulting sensitivity depth curves are such that $\mathcal{D} \propto \sqrt{\Sn} f_0^{-2}$.
For a given ellipticity and distance to the source, low frequencies tend to produce high depth
values (weak signals) as $h_0 \propto f_0^2$ and the PSD of the detector tends ot be very steep.
Towards higher frequencies, where shot-noise is dominant, $h_0$ grows faster than the detector's PSD,
resulting in a monotonically decreasing sensitivity depth (increasing signal amplitude).
For a given ellipticity and distance to the source, a CW signal is ``easier'' to detect at a high frequency
than at a low frequency.

We note that all-sky CW search upper limits in O3 data~\cite{KAGRA:2022dwb}
rule out this optimistic population. Said upper limits, however, are only valid for \emph{deterministic}
CW signals whose frequency evolution is not affected by stochastic effects,
such as glitches~\cite{Ashton:2017wui,Ashton:2018qth} or spin-wandering~\cite{Mukherjee:2017qme}.
For a \emph{generic} CW signal, the results as reported by the \texttt{SOAP} pipeline~\cite{SOAP,SOAPML}
are less constraining. 

\subsection{Visibility of CW signals in a power spectra}
\label{subsec:ifos}

\begin{figure}
    \center
    \includegraphics[width=0.8\textwidth]{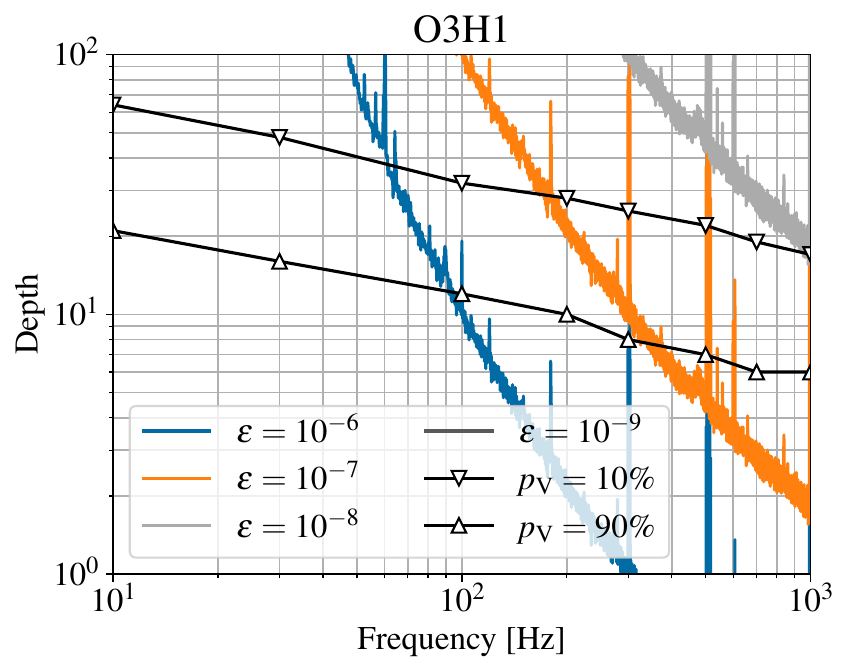}
    \includegraphics[width=0.8\textwidth]{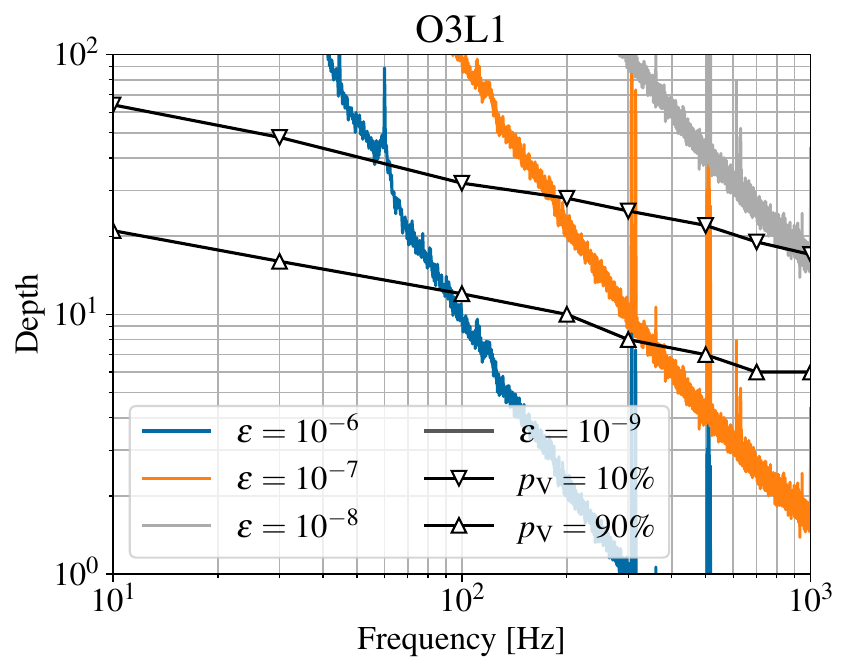}
    \caption{
        Expected sensitivity depth for an optimistic source consistent with a
        NS located at $20\,\mathrm{pc}$ as observed by the
        Advanced LIGO Hanford~\cite{O3bH1ASD} (top)
        and Livingston~\cite{O3bL1ASD} (bottom) detectors during O3.
        Solid lines correspond to different ellipticity values
        ($\varepsilon = 10^{-9}$ is beyond $\mathcal{D} = 100$).
        Note that these are a discretized version of the shaded areas
        in Fig.~\ref{fig:ellipticity_ranges}.
        Triangles correspond to visible sensitivity
        depths at $\pd = 10\%$ and $\pd = 90\%$
        as reported in Table~\ref{table:depths}.
    }
    \label{fig:O3}
\end{figure}

\begin{figure}
    \center
    \includegraphics[width=0.8\textwidth]{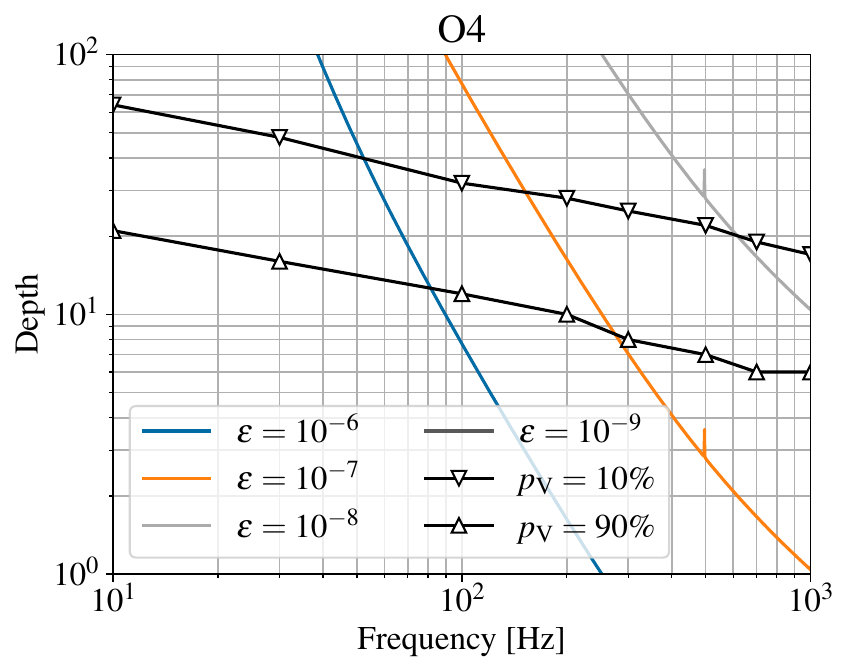}
    \includegraphics[width=0.8\textwidth]{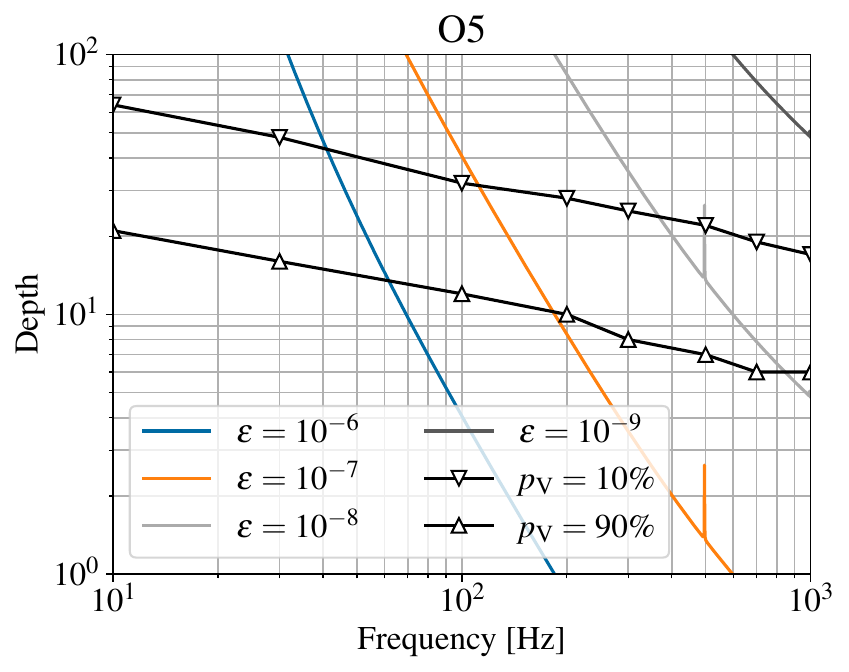}
    \caption{
        Same result as in Fig.~\ref{fig:O3} for the Advanced LIGO
        detectors using the projected sensitivity for
        O4~\cite{O4ASD} (top) and O5~\cite{O5ASD} (bottom).
    }
    \label{fig:O4O5}
\end{figure}

\begin{figure}
    \center
    \includegraphics[width=0.8\textwidth]{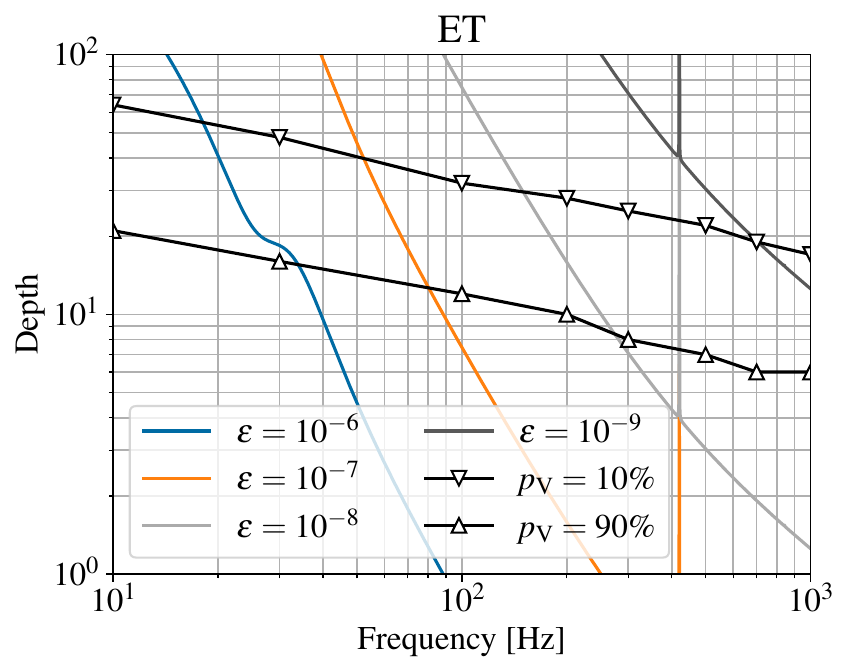}
    \includegraphics[width=0.8\textwidth]{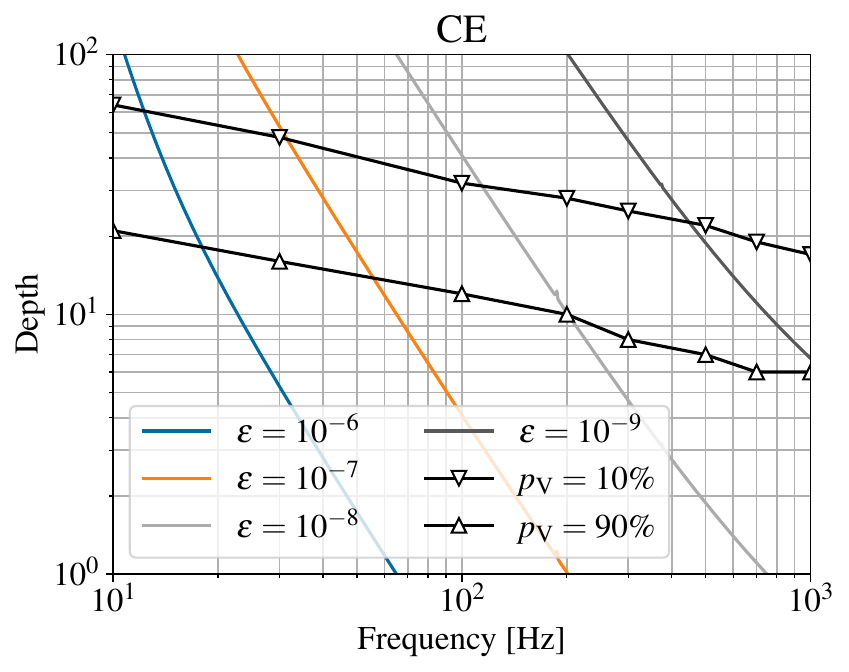}
    \caption{
        Same result as in Fig.~\ref{fig:O3} for the projected 3rd-generation
        detectors ET~\cite{ETASD} (top)
        and CE~\cite{CEASD} (bottom).
    }
    \label{fig:ETCE}
\end{figure}

In Figs.~\ref{fig:O3}, \ref{fig:O4O5}, and \ref{fig:ETCE} we compare the visible
sensitivity depths derived in Sec.~\ref{sec:introduce_kurtosis} to the expected
sensitivity depths from astrophysical CW signals introduced in
Sec.~\ref{subsec:optimistic_cw}. To ease the discussion, we use the results using
$\Tsft = \SI{1800}{\second}$, which are the most conservative. Other $\Tsft$
values produce comparable results as the variations are small.

We are interested in identifying frequency bands for which the visible sensitivity
depth is \emph{higher} than the expected depth for an optimistic CW signal.
In such bands, CW signals will be visible in  the power spectrum.

Overall, we find that visible signals tend to be located in the upper end of the
frequency spectrum. This fact seems to contradict the results from
Sec.~\ref{sec:kurt_from_cw}, where we concluded that high frequencies are
less visible than low frequencies due to a broader Doppler modulation.
The detector's sensitivity curve, however, degrades very steeply towards low
frequencies; towards high frequencies, where shot-noise is dominant,
the frequency dependence is much more gentle compared to $h_0$'s quadratic
dependency on $f_0$, and thus the resulting sensitivity depth quickly becomes
lower enough to be visible. 

For high ellipticity values, $\varepsilon \in  \left[ 10^{-7}, 10^{-6} \right]$,
and for the Advanced LIGO detectors in their O3 configuration
(Fig.~\ref{fig:O3}), CW signals would start to become visible at about
\mbox{$\SI{100}{\hertz}$}. As we progress into O4 and O5 sensitivities
(Fig.~\ref{fig:O4O5}), this frequency may reduce close to \mbox{$\SI{50}{\hertz}$}.
Typical searches for CW signals from unknown sources survey frequencies within
$[20, 1000]\,\unit{\hertz}$~\cite{KAGRA:2022dwb}. As a result, as we approach
the design sensitivity of the Advanced detectors,
CW signals are expected to be visible for $95\%$ of the frequency band.
For 3G detectors (Fig.~\ref{fig:ETCE}), high ellipticity signals will be
visible across practically the whole frequency band.

For low ellipticities $\varepsilon \in \left[ 10^{-9}, 10^{-8} \right]$,
CW signals would become visible starting at $\SI{700}{\hertz}$ during O4
and $\SI{500}{\hertz}$ during O5. For 3G detectors, low ellipticity signals
would be visible from $\SI{100}{\hertz}$ onwards. 

%% file: conclusion.tex
We have studied whether an astrophysical CW signal could be visible as a 
narrow spectral artifact in the power spectrum of a gravitational-wave detector.
To do so, we have calibrated the sample kurtosis of a power spectrum as a measure
of the visibility of a CW signal. We found that, depending on the signal's frequency,
sensitivity depths between $\mathcal{D} = 5$ and  $\mathcal{D} = 60$ could become
visible in a power spectrum, and thus mimic the behavior of a line.

These results were compared to the expected amplitude of an optimistic astrophysical
signal. For high ellipticities $\varepsilon \in \left[10^{-7}, 10^{-6}\right]$,  CW signals
become visible from $\SI{100}{\hertz}$ onwards for the Advanced LIGO detectors
which amount to 95\% of the frequency bands surveyed in a typical all-sky search;
for 3G detectors, signals will be visible across the whole frequency band. For low ellipticities
$\varepsilon \in \left[ 10^{-9}, 10^{-8} \right]$, CW signals would start to become visible
at $\SI{700}{\hertz}$ for the O4 detectors,  $\SI{500}{\hertz}$ for the O5 detectors,
and $\SI{100}{\hertz}$ for 3G detectors. We conclude that it is unsafe to discard CW search
outliers near narrow spectral artifacts of unknown origin. Safely discarding outliers near spectral
artifacts requires evidence that the artifact is instrumental.

Throughout this study we assumed a population of CW signals with a negligible spindown parameter
$f_1 \approx \SI{0}{\hertz / \second}$. Given the typical resolution of an all-sky CW search, this
assumption is consistent with the observed pulsar population~\cite{atnf}. The impact of higher spindown
values would be to spread the power of a CW signal along neighboring frequency bins, reducing
the peak's prominence in a power spectrogram. This is more likely to affect high ellipticity sources,
which are expected to have a higher spindown value. The results here presented tend to overestimate
the visibility of high ellipticity sources.

Also, in this study we neglected two properties of CW signals, namely their distinctive double-horned
shape in a power spectrum~\cite{Valluri:2020cqe} and the fact that they would be expected to show up
in multiple detectors at once. The former requires the CW signal to be located in a relatively quiet frequency
band, as otherwise the expected horns would end up tarnished by instrumental or unknown lines. The latter
requires a network of detectors with comparable sensitivities at a given frequency band, which makes it depend
heavily on the specific configuration of the detectors. 

Our results draw attention to the importance of conducting detector characterization studies
on narrow spectral artifacts, and to the non-negligible possibility of missing an obvious CW signal
if such tasks are neglected. Although a significant (and increasing) number of lines have
been characterized and mitigated thanks to detector characterization efforts within the LIGO-Virgo-KAGRA
collaboration, the origin of a comparable amount still remains, to date, unknown. 
The importance of rigorous investigation to distinguish between instrumental artifacts and genuine CW signals
will only increase as we progress into the era of design sensitivity of the advanced detectors
and further beyond into 3G.